\documentclass[fleqn, usenatbib]{mnras}

\usepackage{newtxtext, newtxmath}
\usepackage{enumerate}
\usepackage{booktabs}

\usepackage[T1]{fontenc}
\usepackage{ae, aecompl}
\usepackage{hyperref}

\usepackage{graphicx}	
\usepackage{amsmath}	
\usepackage{amssymb}	
\usepackage{subfigure}

\title[Spectrum images calibration]{Two-dimensional Multi-fiber Spectrum Image Correction Based on Machine Learning Techniques}

\author[Jiali Xu et al.]{
Jiali Xu,$^{1}$
Qian Yin,$^{1, \star}$
Ping Guo$^{2}$\thanks{E-mail: yinqian@bnu.edu.cn(QY); pguo@bnu.edu.cn (PG)}
and Xin Zheng$^{1}$
\\
$^{1}$Image Processing and Pattern Recognition Lab., 
School of Artificial Intelligence, Beijing Normal University, Beijing 100875, China\\
$^{2}$Image Processing and Pattern Recognition Lab., 
School of Systems Science, Beijing Normal University, Beijing 100875, China\\
}

\date{Accepted XXX. Received YYY; in original form ZZZ}

\pubyear{2020}

\begin{document}
\label{firstpage}
\pagerange{\pageref{firstpage}--\pageref{lastpage}}
\maketitle

\begin{abstract}
Due to limited size and imperfect of the optical components in a  spectrometer, aberration has inevitably been brought into two-dimensional multi-fiber spectrum image in LAMOST, which leads to obvious  spacial variation of the point spread functions (PSFs). Consequently,  if spatial variant PSFs are estimated  directly , the huge storage and intensive computation requirements result in  deconvolutional spectral extraction method become intractable.  In this paper, we proposed a novel method to solve the problem of spatial variation PSF   through  image aberration correction. When CCD  image aberration is corrected, PSF, the convolution kernel,  can be approximated by one spatial invariant PSF only.  
Specifically,  machine learning techniques are adopted to calibrate distorted spectral image, including Total Least Squares (TLS) algorithm, intelligent sampling method, multi-layer feed-forward neural networks.  The calibration experiments on the LAMOST  CCD  images show that the calibration effect of proposed method is effectible. At the same time, the spectrum extraction results before and after calibration are compared,  results  show the  characteristics of the extracted one-dimensional waveform are more close to an ideal optics system,  and the PSF of the corrected object spectrum image estimated by the blind deconvolution method is nearly  central symmetry, which indicates that our proposed  method can   significantly reduce the complexity of spectrum extraction and improve extraction accuracy.
\end{abstract}





\section{Introduction}

With the construction and application of the Large Sky Area Multi-Object Fiber Spectroscopic Telescope (LAMOST), the acquisition efficiency of celestial spectra has been greatly improved\citep{SongLarge, Shi2016The}. But at the same time, researchers need to make more improvements to existing spectrum analysis methods and processing flow dealing with massive data. Two-dimensional (2D) optical fiber spectrum images can be available after the celestial signals being processed by the telescope system and finally imaged on the CCD. 
During the imaging process, the luminous flux (energy) of each fiber at different wavelengths diffuses to adjacent areas in a defined form, forming a fiber profile with distinctive characteristics\citep{ZhangFiber}. The form of diffusion can be described objectively by the point spread function (PSF), which represents the brightness distribution of the output image when the input  is a point light source. One-dimensional spectra, which consists of the actual flux at different wavelengths without the effect of energy diffusion, are supposed to be extracted from 2D spectrum images to help astronomers analyze the physical and chemical properties of celestial bodies. However, the existing spectrum extraction algorithms have their  advantages and disadvantages and cannot guarantee both high accuracy and high efficiency.

\par
\begin{figure}
\label{fig1psf}
\centering
\subfigure{
\begin{minipage}[t]{0.25\linewidth}
\centering
\includegraphics[width=0.8in]{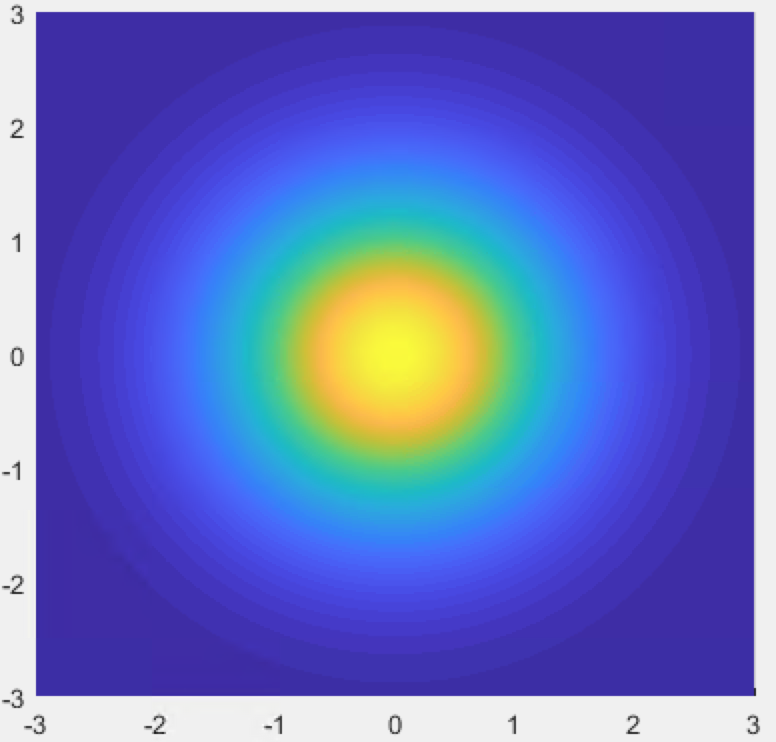}
\end{minipage}%
}%
\subfigure{
\begin{minipage}[t]{0.25\linewidth}
\centering
\includegraphics[width=0.8in]{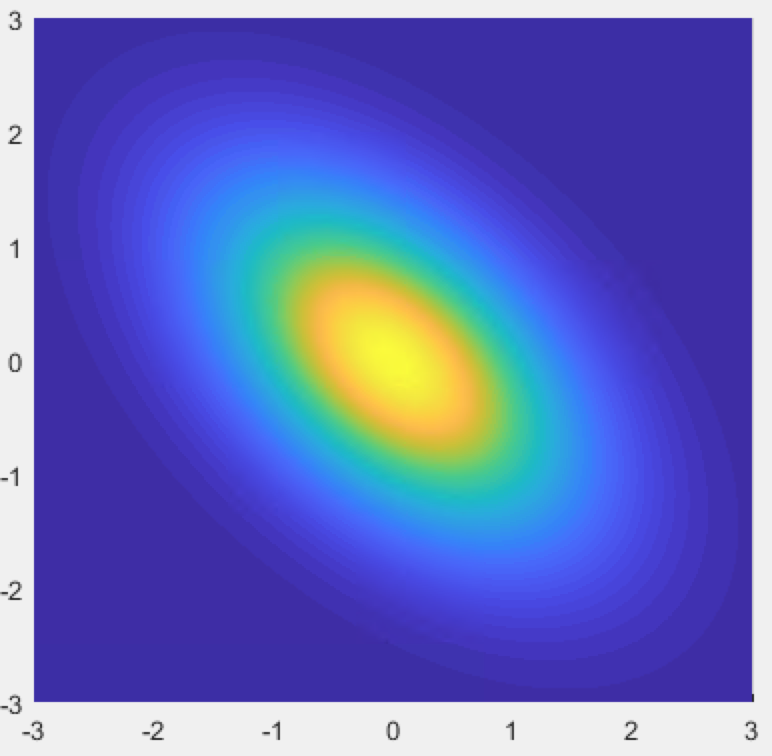}
\end{minipage}%
}%
\subfigure{
\begin{minipage}[t]{0.25\linewidth}
\centering
\includegraphics[width=0.8in]{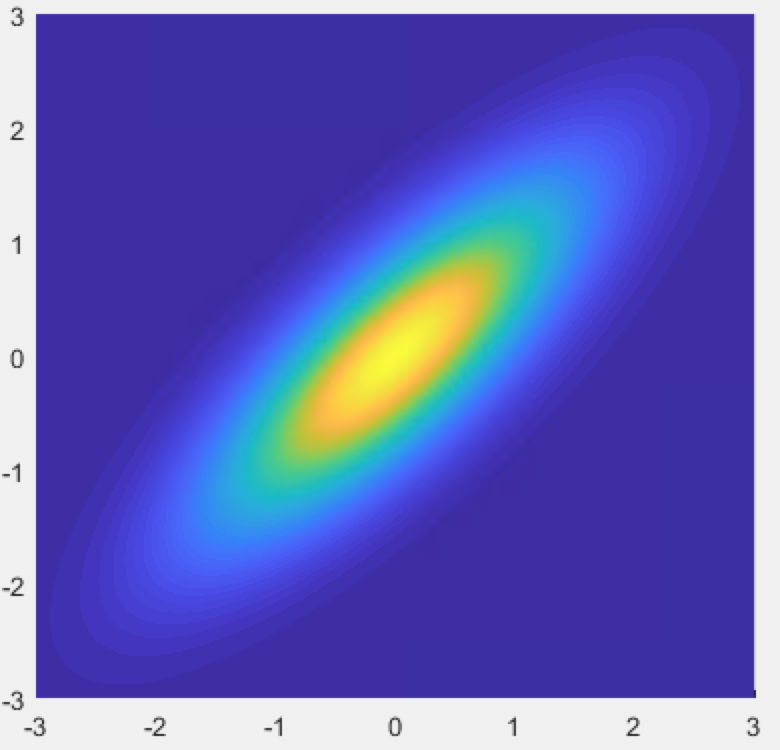}
\end{minipage}%
}%
\caption{Different contours of 2D Gaussian PSF model with different parameters.}
\end{figure}

In commonly used spectrum extraction methods, the aperture extraction method\citep{Boer1982The} and the optimal aperture extraction methods\citep{HorneAn, Robertson1986OPTIMAL} count up the flux within a selected certain radius for each wavelength, the difference between them lies in whether to assign different weights to pixels in the spatial direction. The profile fitting method\citep{2000Msngr, Piskunov2002New, CuiSpectral2009} selects a suitable function to fit the profile of the fiber spectrum in spatial direction. Until 2010, the deconvolution method was applied to spectrum extraction\citep{Bolton2010Spectro}, which is regarded as one of the most accurate and promising methods. The main idea of the deconvolution method is that spectrum extraction can be considered as a direct deconvolution operation when PSF and 2D multifiber spectrum images are known, so it is the only method based on the imaging principle different from the above-mentioned methods, and can characterize the resolution accurately to upper limits of native instruments\citep{Yu2014A, Yin2017Blind}. In the implementation of the deconvolution method, the determination of PSFs is a sustainable challenge. On the one hand, the PSF may vary from time to time and cannot be estimated for every observation night. On the other hand, due to the interference of CCD sensor performance, defocus of the imaging system, atmospheric turbulence effect and other factors, the actual obtained two-dimensional optical fiber spectrum images have noticeable distortion and the contours of PSFs at different spatial locations are diverse from each other.  The spacial variant PSF is manifested in obvious bending in the flat-field images and spots of various shapes in the calibration arc images. At present, several types of analytical expressions for PSFs, such as the 2D Gaussian\citep{ZhuSpectrum, Yu2014A} (Fig. \ref{fig1psf} shows some examples of Gaussian type PSF. )  and 2D exponential polynomial\citep{ZhangFiber} is used to model PSFs, and the flat-field spectrum and the calibration lamp spectrum are used to determine the parameters of the function in most studies. But the PSF estimation method is not practical to integrate into the LAMOST data processing pipeline because it requires interactively selecting the PSF form and estimating the model parameters. Considering the difficulty in determining PSFs of large-scale multi-fiber spectral telescope in the actual environment, \cite{Yin2017Blind} used the blind deconvolution extraction method to extract one-dimensional spectrum when the PSF of the imaging system was unknown, which improved the applicability of deconvolution decimation to some extent. However, most available extraction methods, including the blind deconvolution method, do not really solve the problem that the shape and range of the PSF varies with space and wavelength. In order to reduce computational complexity, each fiber uses the same PSF.  Actually, many relevant studies have discussed the estimation of spatial variant PSF in the field of image restoration. \cite{Sawchuk1972Space} used the spatial coordinate transformation method to reconstruct PSF images of spatial changes. The idea of this method is to find a coordinate space, and the degradation of each point of the image after transformation to this space is spatially invariant. After the restoration of this space by general algorithm, the image will be converted back to the original image space. The premise of the spatial coordinate transformation method is that the spatial variation law of the degenerate function can be expressed analytically \citep{Sawchuk1973Space, Sawchuk1974Space}.\cite{Trussell1978Image} proposed a restoration algorithm based on image segmentation, which was further developed by \cite{Costello2003Efficient}. The important reason that these existing methods are not applied to optical fiber spectroscopy is that the computational complexity is too high to meet the requirements of massive large-scale spectral image processing, especially for a telescope like LAMOST, 250 optical fibers are arranged on a CCD image of $4096\times 4096$ pixels at the same time.

\par
Considering that ignoring the spatially varying PSFs will have a great impact on the final spectrum result and the computational complexity of directly modeling the PSFs is extremely too high to achieve, we propose a preprocessing method by spectrum calibration to eliminate the inconsistency of PSFs at different positions. Ideally, the PSF should be symmetric with a center circle and the PSF at each point in the image should be consistent. With such processing of image calibration, the difficult task of modeling spatially varying PSFs can be avoided and we can achieve the purpose of reducing the complexity and ensuring the accuracy of the spectrum extraction. For the task of spectrum calibration, \cite{Zhu2019Bending} proposed a normal mapping method to correct the bending of the 2D optical fiber spectrogram. Each fiber spectrum in the two-fiber spectrogram is processed separately without considering the correlation between different fibers. Due to the complexity of the spectral image imaging process, we think that artificially designing the transformation relationship is not the best solution. In contrast, the spectrum calibration method based on machine learning techniques proposed in this paper has better practicability and accuracy.

\par
The reminder of the paper is organized as follows. Section~\ref{sec:background} describes the background on  spectrum distortion, and the proposed spectrum calibration frameowrk is described in Section~\ref{sec:techniques}. A discussion of our experiments and results is provided in Section~\ref{sec:results}. Our conclusion is presented in the last section.

 \section{spectrum image distortion}
 \label{sec:background}
 
 \begin{figure}
    \centering
    \includegraphics[width=7.0cm]{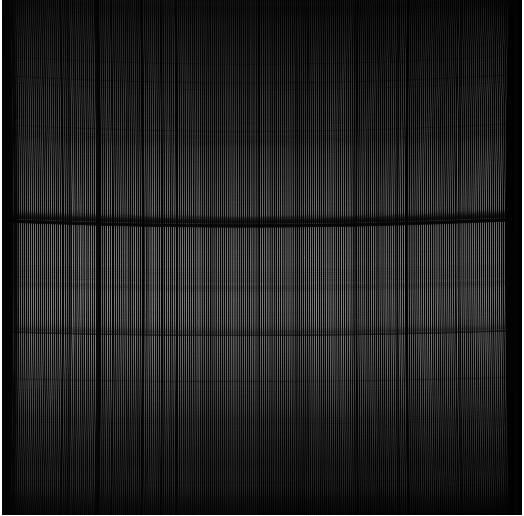}
    \caption{Original LAMOST flat-field spectrum image. Optical fibers imaging lines have different degree of distortion in this image.}
    \label{fig:flat}
\end{figure}
Firstly, let us consider the principle of deconvolution spectrum extraction. The optical instruments, such as collimator mirror, dichroic beam splitter and diffraction grating, will disperse the detected light of astronomical objects and give rise to broadening and effects on the measured signals. The observed image is obtained by the convolution of the source image and a PSF,  which can be represented generally by the following integral equation,
  \begin{equation}
  g(x,y)=\iint_{-\infty }^{+\infty}h(\xi  - x, \eta -y)f(\xi , \eta)d\xi  d\eta + \varepsilon ,  
\end{equation}  
where $g(x,y)$ represents the image intensity at the position $(x, y)$ in the observation plane,
$f(\xi , \eta)$ represents the light source at position $(\xi , \eta)$ , $h (\xi  - x, \eta -y)$ represents the PSF value, and $\varepsilon$ denotes an additive noise at the position $(x, y)$. This means that the process of spectrum extraction can be viewed as a deconvolution operation for solving $f$, given $g$ and $h$ are known.  

\par
 According to the imaging principle, the process of spectrum extraction is actually the operation of deconvolution.  
The deconvolution method, which is  often used in image restoration, was not applied to spectrum extraction until 2010 \citep{Bolton2010Spectro}. But in their experiments, analog data instead of large-scale actual observation data is used to prove the validity of this algorithm, meanwhile, they also  pointed out computing challenges posed by convolution operation on massive data was inevitable and recommended to perform parallel data processing using computer's GPUs. In practice, another factor limiting the further application and higher accuracy of the deconvolution method is the estimation of the PSF, especially for large spectroscopic telescope like LAMOST. In order to cover a larger range of scenery,  wide-angle lens is widely used in image shooting. When we get benefits from the large field of view, it is inevitable to bring the distortion into the imaging process while the proportion of distance is not coordinated. Although this kind of image distortion does not affect the resolution of the image, it will reduce the geometric precision of the object in the image, so as to bring about the error. In addition, the degree of image distortion will increase with the increasing field of view, at the mean time the image quality will become worse as well. This wide-angle distortion combined with other complex effects, such as  CCD sensor performance, imaging system defocus and atmospheric turbulence, makes us have to face a challenge -- spacial variant PSF. When celestial objects are imaged on two-dimensional optical fiber spectra, location, shape and scope of PSF is changing with the variation of space position, which is due to aberration in the spectrometer and asynchronous vibration of various components. However, due to the high complexity of space-variant PSF modeling and deconvolution calculation after modeling, it is almost impossible to meet the requirements of large-scale spectral data processing. 

\par
\begin{figure}
    \centering
    \includegraphics[width=0.6\columnwidth]{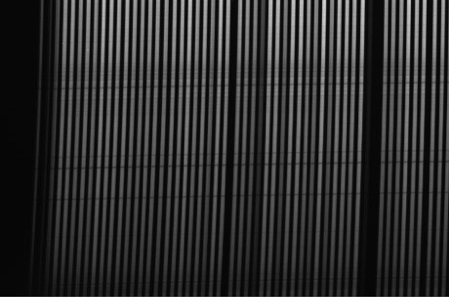}
    \caption{Part of the spectral image with obvious distortion.}
    \label{fig:local}
\end{figure}

Faced with such a real teaser, we can choose another shortcut, which is to eliminate distortion by correcting the original image, so that the PSFs at different spatial positions tend to be consistent and we can avoid complex PSF modeling. In the next section, we will propose a specific spectrum calibration framework for how to eliminate this nonlinear distortion. At present, what we need to make clear is the expected corrected image. Fig.\ref{fig:flat} is a flat-field spectrum of LAMOST showing visible pillow type of distortion. The closer to the edge, the greater distortions are.  Obvious distortion nearing the edges can be seen in Fig.\ref{fig:local}. 
In the absence of optical distortion of the spectrometer and CCD camera,  PSF appears as a symmetrical round outline and keeps same at each point while all the spectra are expected to be straight, parallel to each other and neatly aligned on the CCD under ideal conditions. If we can correct these bent spectra to be straight, the accuracy of spectrum extraction will be greatly improved while the difficulty will be reduced a lot, and the system's data analysis and processing capabilities can be greatly improved. Therefore, there is an urgent demand to calibrate the non-linear distortion of the spectrum.

\section{Spectrum image correction  with machine learning techniques}
\label{sec:techniques}

 \begin{figure*}
    \centering
    \includegraphics[width=140mm]{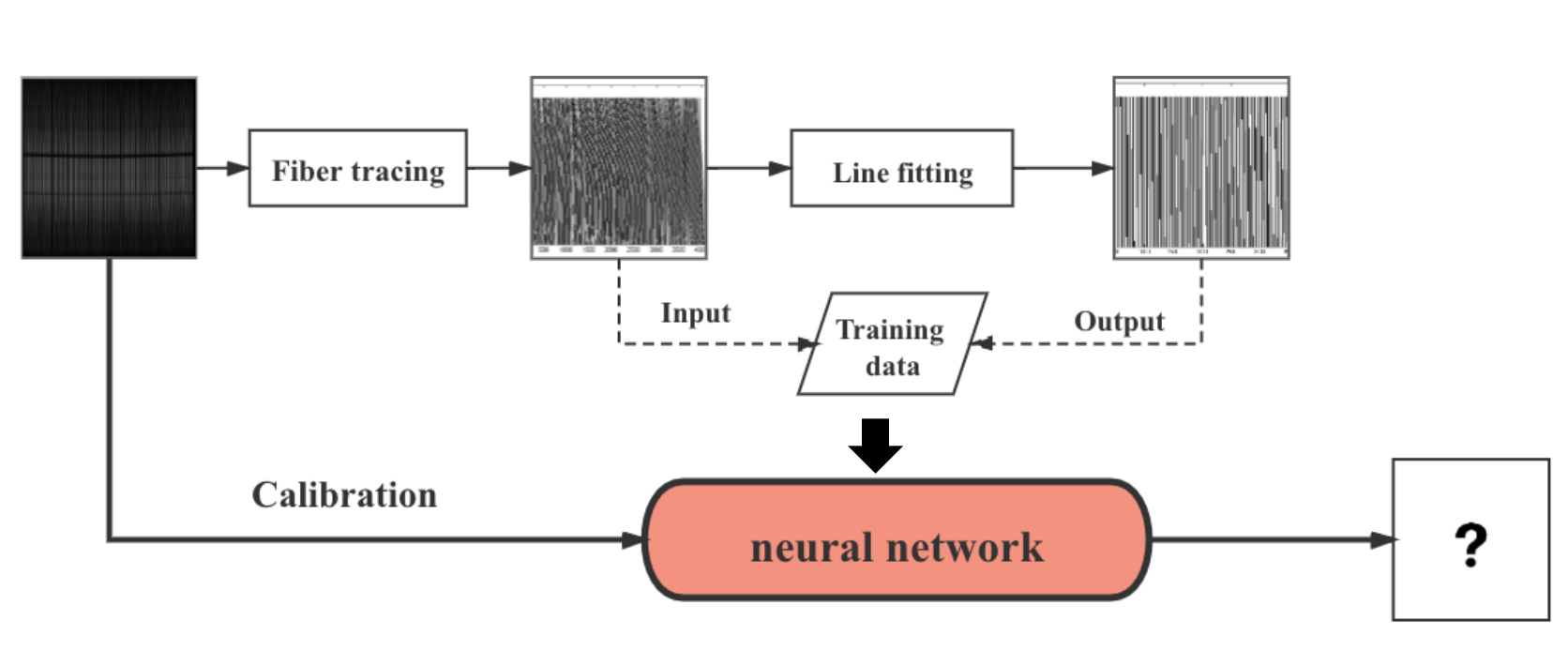}
    \caption{Spectrum image correction method based on neural network.}
    \label{fig:method}
\end{figure*}
\label{sec:method}
 \begin{figure}
    \centering
    \includegraphics[width = \columnwidth]{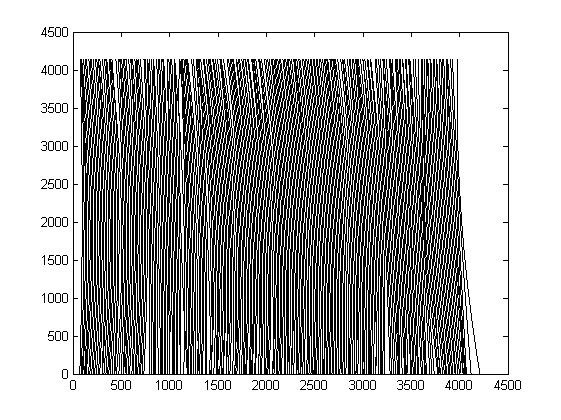}
    \caption{Fiber tracing result.}
    \label{fig:tracing}
\end{figure}

In general, establishing a corresponding mathematical model according to the reason of image distortion and correcting the image according to the inverse process of imaging is a basic idea of  image correction. Building an accurate mathematical model is the key to obtaining excellent image calibration results. However, in the real LAMOST environment, the process of spectrum imaging on the CCD is disturbed by various factors, such as the comprehensive action of CCD sensor performance, imaging system defocus and atmospheric turbulence. This distortion brings a lot of error to the subsequent data processing processes like spectrum extraction and flux calibration. It is extremely difficult to describe the interference by explicitly constructing nonlinear functions artificially. As we all know, the neural network can model complex function through nonlinear mapping of activation function, the combination of multiple neurons and a large number of sample training, showing strong self-learning ability and self-adaptability. \cite{Hornik1989Multilayer} showed in their works that feed-forward neural networks are indeed capable of universal approximation in a very precise and satisfactory sense, even with as few as a single hidden layer. Inspired by this, we propose to apply the multi-layer neural network in automatically generating a complex transformation model between distorted and non-distorted images. The entire spectrum correction framework is shown in the Fig.\ref{fig:method}. The training of the neural network is a nonlinear optimization problem, looking for a combination of parameters to minimize the objective function. The whole implementation process needs a large number of training samples for the supervised learning task. Unlike common image classification and detection tasks, we do not own ideal spectrum images without distortion, and manually labeling  output samples is even more infeasible. For spectrum image correction, the construction of input and output sample pairs and the selection of proper number of samples are the primary problems that need to be solved, which is also the difficulty and innovation of our proposed method should be considered \citep{li2017ijcnn}.

\subsection{Input construction by fiber tracing}
First we give a simplified description of the calibration problem. The essence of the task is to find a transformation relationship to rearrange the pixels in the image, that is, the pixels with the original coordinates of $(x, y)$ are transformed into $(x', y')$, as shown in equation~(\ref{eq:transform}). The calibration of the LAMOST two-dimensional spectrum is a kind of nonlinear distortion calibration with large field of view, including radial distortion and tangential distortion \citep{Lin2005Distortion}, in which the tangential distortion can be almost ignored \citep{Tsai1922Aversatile}. In optical imaging, a straight line through the optical center is still linear in the imaging process, and the minimum distortion in the optical axis nears the center. 

\begin{equation}
I \left ( x  , y \right ) \overset{f(x, y)} {\Longrightarrow}  I \left ( {x}' ,  {y}' \right). 
\label{eq:transform}
\end{equation}

\par
Fiber tracing is one of the common pre-processing steps for spectrum extraction, which can obtain the center trace of 250 spectra from the two-dimensional spectral image. Fig.\ref{fig:tracing} shows the fiber tracing result of the 250 spectra in a spectral image. Every fiber trace is a curve passing through the center of the spectral profile, generally expressed by the low order polynomial, such as Legendre Polynomials. Apparently, the tracing result can well reflect the distortion characteristics of different spectra. Different fiber trace in the spatial direction are different, and the curve bending near the edge is more obvious. Therefore, coordinates of points on the curve are fed into neural networks as input vector in our method. The input vector includes two attributes, one attribute indicates the pixel index in the wavelength orientation, and the other attribute indicates the pixel index in the spatial orientation.

\par
\subsection{Generating labels with the Total Least Square method}
The neural network is essentially a nonlinear optimization problem, which is looking for a set of parameters combination to minimize the objective function according to the known constraints. To supervised train a feed forward  neural network, the construction of training samples is of great importance. The spectral center curves before correction through fiber tracing has been obtained, but we only got the input of the training samples, and the output is unknown. Based on the prior knowledge that optical fiber spectra are lines parallel to each other ideally, the corrected spectral center should be a straight line. Therefore, we use the line fitting method to automatically construct the corresponding output. Constructing labels based on experience is also a common strategy for unlabeled data in machine learning. In order to get the best fitting effect, the primary problem to be solved is the choice of a straight-line fitting model. One of the most common fitting criteria is the classic Least Squares (LS) method, but the LS method assumes that all errors are derived from the dependent variables and ignores the errors of the independent variables. In our proposed methodology, we utilize the Total Least Square (TLS) method to carry out the curve fitting. Compared with the LS method, the TLS method take into account the error of both independent variables and dependent variables\citep{de1998TSL}, which can help improve the accuracy of line fitting. The fitting lines are very close to the imaging results without distortion, so we can approximately think that it is just the corrected target output. 

\par
In order to better understand the TLS method, we first introduce the traditional LS method and then explain the idea of TLS on this basis. For some discrete data points $(x_{i},y_{i}) (i=1,2,3,...,m)$ passing through the same line, they all meet the general form of a line,
\begin{equation}
    y_{i}=ax_{i}+b ,
\end{equation}
where $a$ is the slope of the line, and $b$ is the intercept. $a$ and $b$ are the parameters to be estimated. Although there is no guarantee that all points are gathered in a straight line, it is expected that these points are located in the vicinity of the line as far as possible. Let $a_{0}$ and $b_{0}$ be their approximations, satisfying $a=a_{0}+\Delta a$,  $b=b_{0}+\Delta b$.

With $y$ as the dependent variable and $x$ as the independent variable, the error equation is expressed as,
\begin{equation}
\boldsymbol {v}_{yi}=\begin{bmatrix}
x_{i} & 1
\end{bmatrix}
\begin{bmatrix}
\Delta a\\ 
\Delta b
\end{bmatrix}+(a_{0}x_{i}+b_{0}-y_{i}) .
\end{equation}

The matrix expression of the error equation is,
\begin{equation}
\mathbf{A} (\Delta \mathbf{X}) = \boldsymbol{L}+\mathbf{V}, 
\end{equation}
where $\mathbf{A}=\begin{bmatrix}
x_{1}  & 1\\ 
x_{2} & 1\\ 
\vdots & \vdots\\ 
 x_{m} & 1 
\end{bmatrix}$, 
$\boldsymbol{L}=\begin{bmatrix}
a_{0}x_{1}+b_{0}-y_{1}\\ 
a_{0}x_{2}+b_{0}-y_{2}\\ 
\vdots \\ 
a_{0}x_{m}+b_{0}-y_{m}
\end{bmatrix}$ , 

$\mathbf{V}=\begin{bmatrix}
v_{y1}\\ 
v_{y2}\\ 
\vdots\\ 
v_{ym}
\end{bmatrix}$, $\Delta \mathbf{X} = \begin{bmatrix}
\Delta a\\ 
\Delta b
\end{bmatrix}$ .

According to the least squares principle, the  $\mathbf{V^{T}}\mathbf{V}$ should be minimized,
\begin{equation}
\text{Minimizing } \sum_{i=1}^{m} \left \| ax_{i}-b-y_{i} \right \|^{2}.
\end{equation}

It will lead to  least squares solution,
\begin{equation}
\Delta \mathbf{\hat{X}} = (\mathbf{A^{T}A})^{-1}\mathbf{A^{T}}\boldsymbol{L}.
\end{equation}

The traditional least square method considers that the measurement error only exists in the dependent variable $y$, but observational errors on both dependent and independent variables are taken into account in TLS, so the linear equation should be redefined as,
\begin{equation}
    y_{i} + v_{yi}=a(x_{i}+v_{xi})+b .
\end{equation}
In the total least square method,  both $\boldsymbol{L}$ and the coefficient matrix $\mathbf{A}$ have errors, so the error equation can be described as,
\begin{equation}
    (\mathbf{A}+\mathbf{E}_{A})\mathbf{X} = \boldsymbol{L} + \mathbf{E}_{l} ,
\end{equation}
where $\mathbf{E}_{A}$ and $\mathbf{E}_{l}$ represent the errors of matrix $\mathbf{A}$ and observation vector $\boldsymbol{L}$ respectively.

Construct the augmented matrix 
$\mathbf{C}= [ \mathbf{A}\,\,\,\,\, \boldsymbol{L}]$, 
and perform QR decomposition $\mathbf{C}=\mathbf{QR}$,
\begin{equation}
    \mathbf{R}=\mathbf{Q^{T} C}=  \mathbf{Q^{T}} \left [ \mathbf{A}_{1}\,\,\,\ \mathbf{A}_{2}\,\,\, \boldsymbol{L} \right ]=
\left[\begin{array}{c c c}R_{11} & R_{12}  & R_{1l}\\0 & R_{22} &  R_{2l} \end{array}\right] ,
\end{equation}
where $\mathbf{A}_{1} = 
\begin{bmatrix}
1 & 1 & ... & 1 
\end{bmatrix}^{T}$, 
$\mathbf{A}_{2} = 
\begin{bmatrix}
x_{1} & x_{2} & ... & x_{m} 
\end{bmatrix}^{T}$.

Construct the augmented matrix $\mathbf{C}_{R}= \begin{bmatrix}
R_{22} & R_{2l}
\end{bmatrix}$,  and perform singular value decomposition,
\begin{equation}
 \mathbf{C}_{R}= \mathbf {U\Sigma N^{T}}, 
\end{equation}
where $\mathbf{\Sigma} = \text{diag} (\sigma _{1}, \sigma _{2}), \sigma _{1}> \sigma _{2}$.

The solution of the TLS is,
\begin{equation}
\begin{bmatrix}
 \Delta \hat{a}\\
\Delta \hat{b}
\end{bmatrix} = \begin{pmatrix}
 \mathbf{A^{T}A}-\sigma _{2}^{2}\begin{bmatrix}
0 & 0\\ 
0 & 1 
\end{bmatrix}
\end{pmatrix}^{-1} \mathbf{A^{T}}\boldsymbol{L}.
\end{equation}

\begin{figure}
    \centering
    \includegraphics[width=\columnwidth]{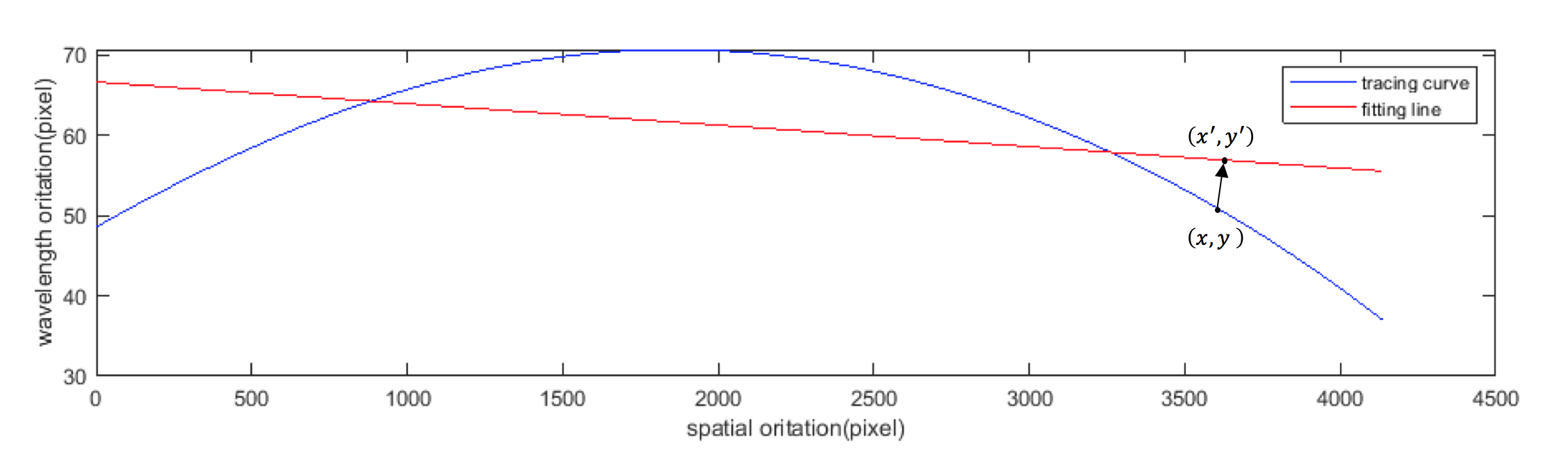}
    \caption{TLS fitting result of one spectrum line.}
    \label{fig:lines fitting}
\end{figure}

The more specific solution process has been described in detail in elsewhere\citep{de1998TSL, MarkovskyOverview, Ding2010}. Take one spectrum as an example, as shown in the Fig. ~\ref{fig:lines fitting},  the dotted line represents the tracing curve of a spectrum, and the other solid straight line represents the result fitted by the TLS method. For convenience, we set the x-coordinate represents the wavelength and the y-coordinate represents the spatial position. It should be noted that the point $(x', y')$ after calibration corresponding to the point $(x, y)$ on the original tracing curve should be the feet on the fitted straight line.

\subsection{Non-uniform sampling}
\label{sec:selection}
\begin{figure}
    \centering
    \includegraphics[width=0.9\columnwidth]{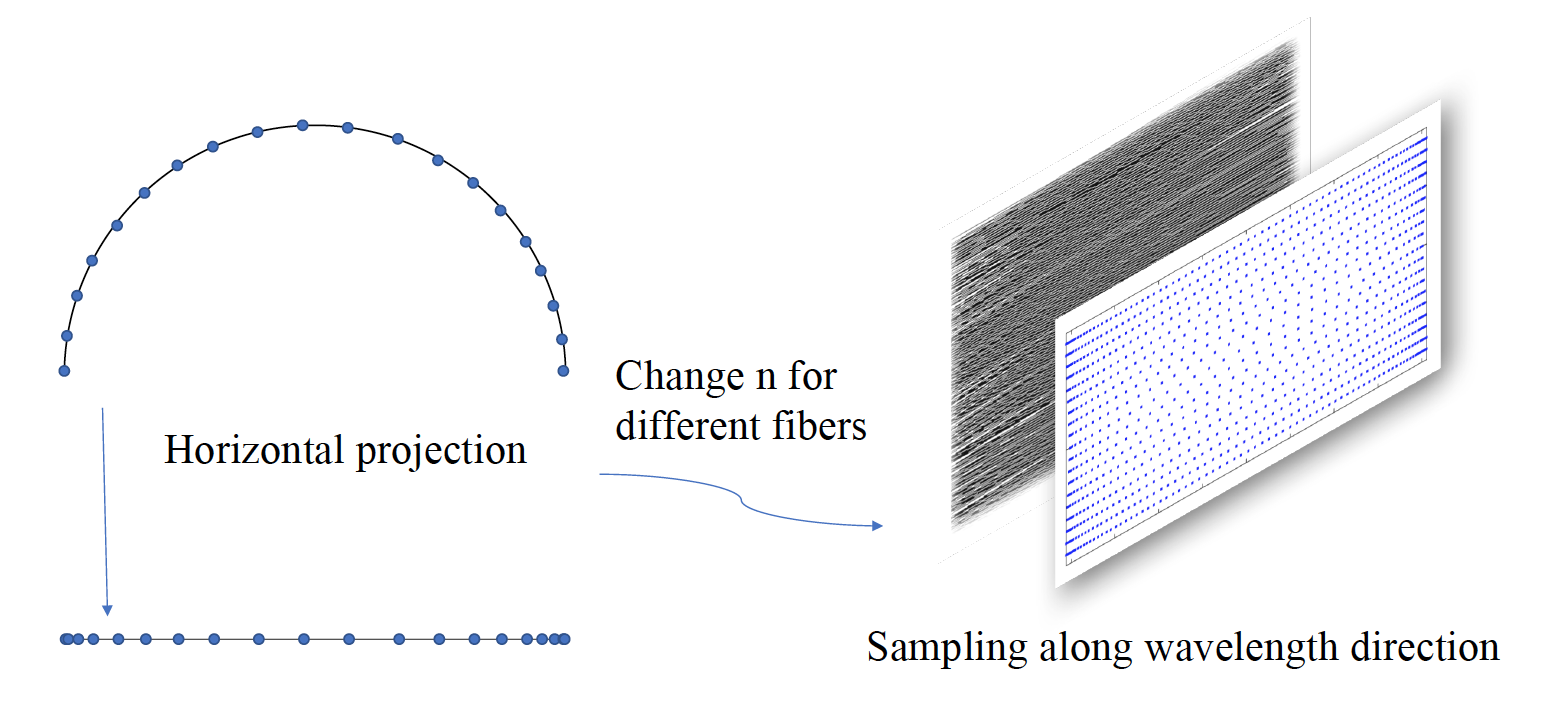}
    \caption{Non-uniform sampling method illustration: samples are obtained through  sample is obtained by projecting uniform distributed points on the arc to the straight line.}
    \label{fig:circle}
\end{figure}

As mentioned before, the distortion of the LAMOST two-dimensional spectral image is mainly a pincushion distortion of a wide-angle lens. This nonlinear distortion is concentrated around the image, and the distortion in the image center is not obvious. Moreover, the size of each spectral image of LAMOST is relatively large, usually $4096\times 4096$, with 250 optical fibers arranged on it. After the fiber tracing, the one-dimensional vector with a size of 4096 can be extracted from each spectrum, which means that a total of $250\times 4096$ pixels will be obtained. The millions of pixels in each image do not need to be all used for training the neural network, because a large part of the data provides redundant information. For example, with no distortion or is not obvious distortion, the pixels in the middle part can be represented by a smaller number of samples. Effectively selecting key samples has a great impact on the final image calibration result. To fully extract the transformation relationship to distortion calibration while ensuring high training efficiency, we designed a non-uniform sampling method to control samples dense in the surrounding area and sparse in the middle. 
This intelligent sampling method can reduce the computation complexity without loss network correction accuracy.  The non-uniform sampling is reflected in two aspects. On the one hand, for each spectrum, the sampling points are non-uniformly distributed along the wavelength orientation. On the other hand, the number of sampling points is different for different spectra along the spatial orientation, specifically reflected as fewer samples in the middle and more samples in the surrounding area. The specific description of the sampling method is as follows.

\par
First,  $N$ points of equal arc length are distributed on the semicircle arc, and all points are projected on the horizontal axis to obtain a one-dimensional vector. At this point, the points in the middle are sparsely distributed and become denser they get closer to the two ends. This one-dimensional vector can be expressed as,

\begin{equation}
\begin{aligned}
      \boldsymbol{x}&=(x_{1},x_{2},...,x_{i},...,x_{n}) \\
      &= (0,\text{R}-\text{R}cos(\frac{\pi}{n-1}),..., \text{R}-\text{R}cos(\frac{\pi * (i-1)}{n-1}),...,2\text{R}) ,
\end{aligned}
\label{eq:R}
\end{equation}
where $R$ represents the radius of the arc. For LAMOST, $R$ can be taken as 2048. 

\par
For each trace curve, increase the value of n from the middle to both sides appropriately, calculate coordinates in the wavelength direction according to equation (\ref{eq:R}) and sample by rounding.
Repeatedly select sample points on the 250 tracking curves to finalize the final training set, and each sample contains two values representing the pixel index in the spatial direction and the wavelength direction, respectively. The overall sample distribution of selected points is shown in Fig.\ref{fig:circle}.\par

\section{Experiments and Discussion}
\label{sec:results}

In the experiment, dataset  from the National Astronomical Observatories,
Chinese Academy of Sciences was used to perform spectral image correction processing\footnote{The dataset  can  be downloaded  from the web site: 
http://dr3.lamost.org/.}.  Among the several kinds of raw two-dimensional data obtained by the LAMOST observation system, the spectral energy of the flat-field spectrum image is distributed uniformly over the entire band, and the spectral profile is relatively smooth, so it is usually used for optical fiber tracing to obtain center curves of multiple spectra. As shown in Fig.\ref{fig:flat}, the resolution of the flat-field spectrum image is $4096 \times 4096$ pixels, the abscissa represents the spatial orientation, the ordinate represents the wavelength orientation, and the magnitude of the gray value represents the flow value at that point, proportional to the number of photons detected during the exposure by each CCD pixel \citep{HorneAn}.

\subsection{Implementation of spectrum image correction}
\begin{figure}
    \centering
    \includegraphics[width = \columnwidth]{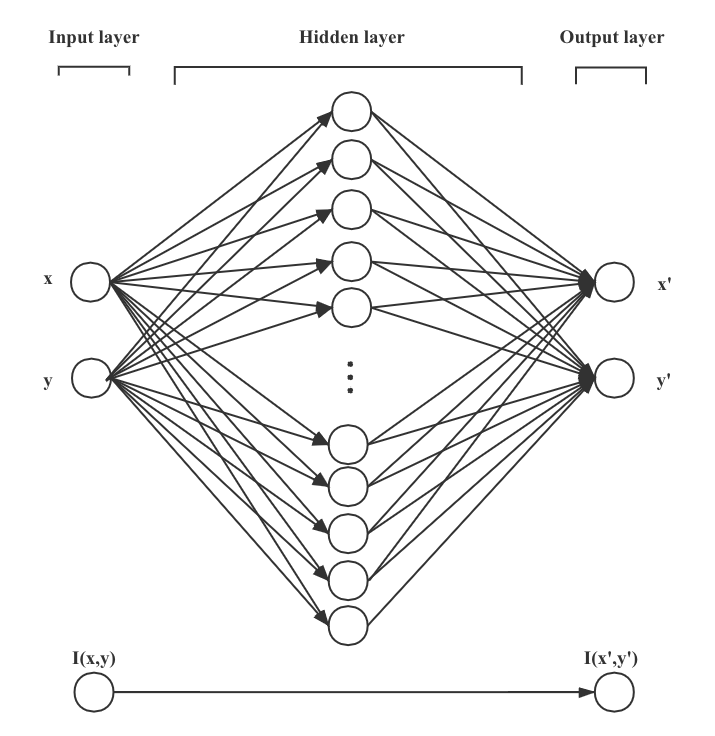}
    \caption{A sketch diagram of the single hidden layer neural network model.}
    \label{fig:nn model}
\end{figure}

\begin{table}
	\centering
	\caption{Neural network training and parameters setting.}
	\begin{tabular}{l l}
		\toprule  
	Hidden layer number  &	1 \\
	Neurons  & (3,20,3) \\
	Activation function & ReLU \\
	Loss function & MSE \\
	Batch size &  600 \\
	Epoches &   50 \\
	Iterations time & 10000 \\
	Optimizer  & Adam \citep{kingma2014adam}\\ 
		\bottomrule  
	\end{tabular}
	\label{tb:parameter}
\end{table}

As described in Section~\ref{sec:method}, in the calibration of two-dimensional spectral images, we need a suitable training set to determine the parameters of the distortion calibration model. The center curves obtained by optical fiber traces can perfectly reflect the distortion of different spectra, and it is not difficult to find that they are the best materials for providing training samples. Fig.\ref{fig:tracing} shows the result of fiber tracing of 250 spectra. It can be seen that the points in different wavelength positions on the same curve have large spatial differences. We selected 30,000 discrete points as input samples of the neural network according to the non-uniform sampling method introduced in Section~\ref{sec:selection}. Each input sample can be represented as a two-dimensional vector $(x, y)^{T}$. Then, 250 center curves in the tracing result were fitted with the TLS method, and the output samples were the feet of the original curve samples on the fitted straight line.\par
In the mathematical theory of artificial neural networks, the universal approximation theorem\citep{Hornik1989Multilayer} shows that a feedforward neural network can fit a function of arbitrary complexity with any accuracy as long as it has a single hidden layer and a limited number of neural units. So in this paper, we design a single-layer neural network to correct the spectrum image, the sketch diagram of the network structure is shown in Fig.\ref{fig:nn model}.  Table~\ref{tb:parameter}
shows the specific parameters setting, such as detailed topology of the network, transfer function,  the chosen performance function and parameters in the training process of the neural network. The fast pseudoinverse  learning algorithm is also utilized to training the neural network \citep{guo1995exact, guo2001pseudoinverse, guo2004pseudoinverse}, and better generalization performance is obtained compared with stochastic gradient descent optimization algorithm.  After training, the coordinate points in the spectral image are fed into the trained network as inputs for testing. The width of each spectrum is about 15 pixels. The coordinates of the 7 pixels on the center and both sides of the spectrum are fed into the trained network as inputs for testing, and the flux value of the corresponding point of the output coordinate is equal to the flux value of the input point, while the other pixels keep the previous flux value.

\subsection{Analysis of results}
\begin{figure*}
\centering
\subfigure[Before  correction]{
\begin{minipage}[t]{0.8\columnwidth}
\centering
\includegraphics[width=\columnwidth]{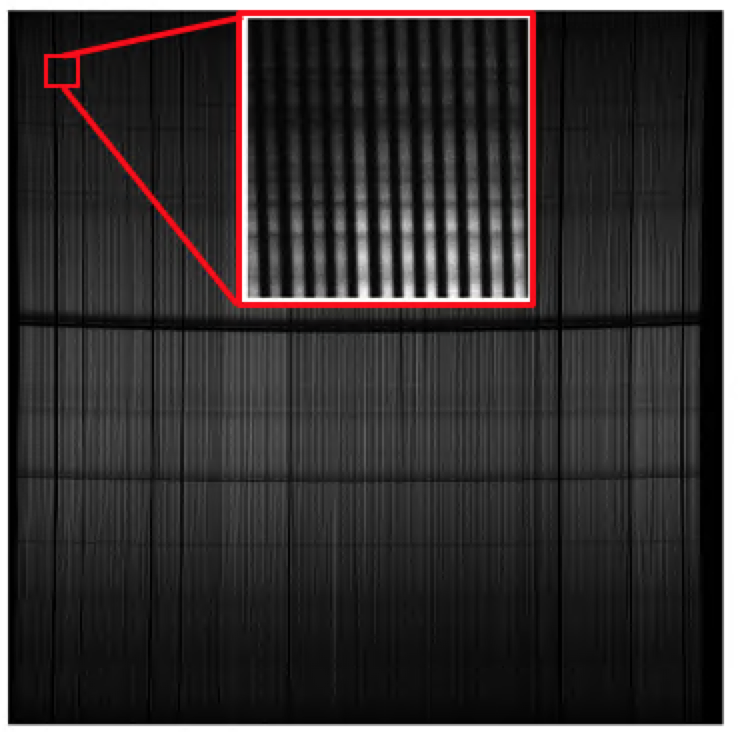}
\end{minipage}%
}%
\subfigure[After correction]{
\begin{minipage}[t]{0.8\columnwidth}
\centering
\includegraphics[width=\columnwidth]{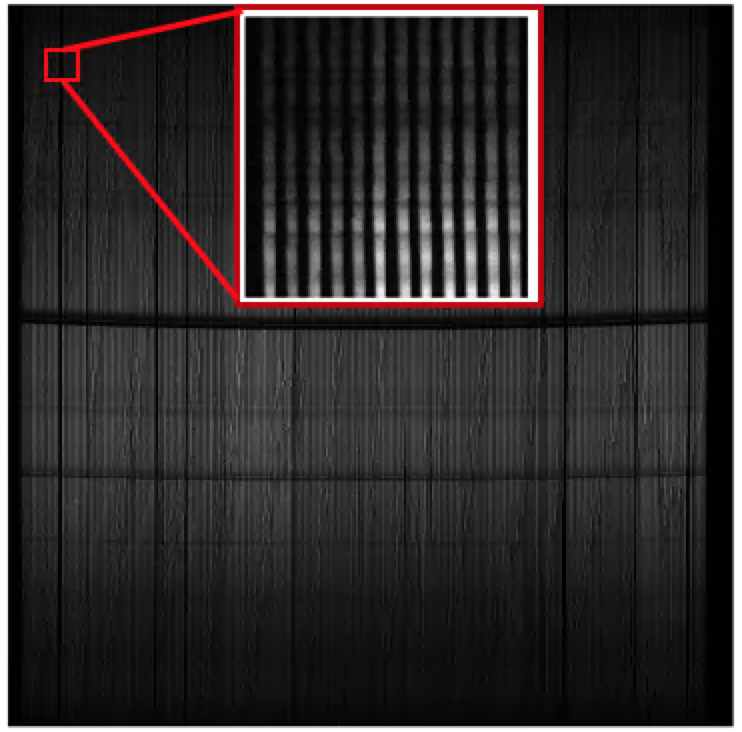}
\end{minipage}%
}%
\centering
\caption{Flat-field spectrum comparison before and after correction.}
\label{fig:comp}
\end{figure*}

\subsubsection{Flat-field spectrum}
\begin{figure*}
\label{fig:traceCompare}
\centering
\subfigure[Before  correction]{
\begin{minipage}[t]{0.8\columnwidth}
\centering
\includegraphics[width=\columnwidth]{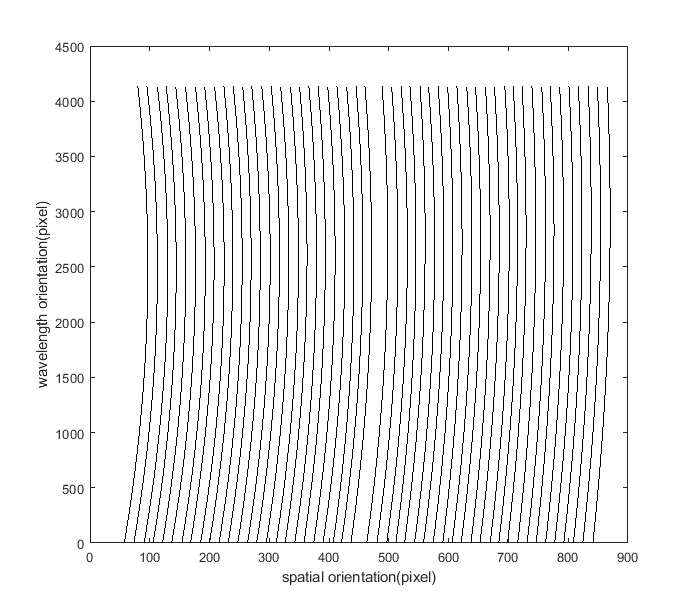}
\end{minipage}%
}%
\subfigure [After correction]{
\begin{minipage}[t]{0.8\columnwidth}
\centering
\includegraphics[width=\columnwidth]{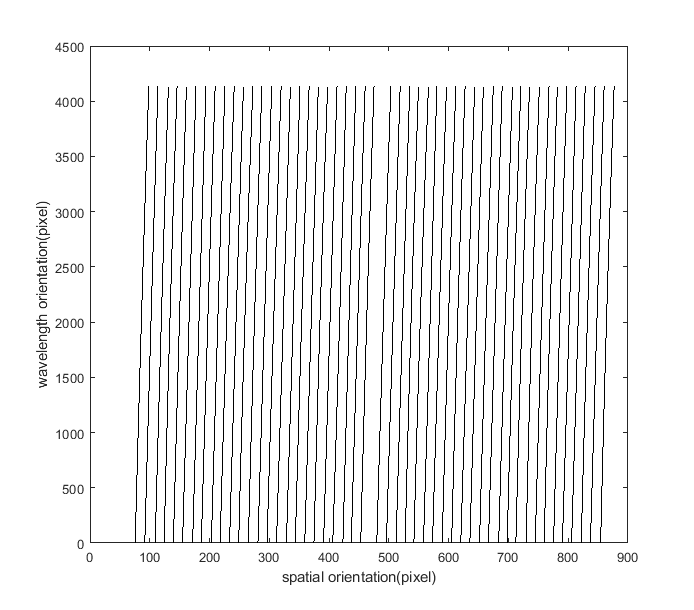}
\end{minipage}%
}%
\centering
\caption{Fiber tracing comparison before and after correction.}
\label{fig:after_tracing}
\end{figure*}

  \begin{figure*}
    \centering
    \includegraphics[width = \linewidth]{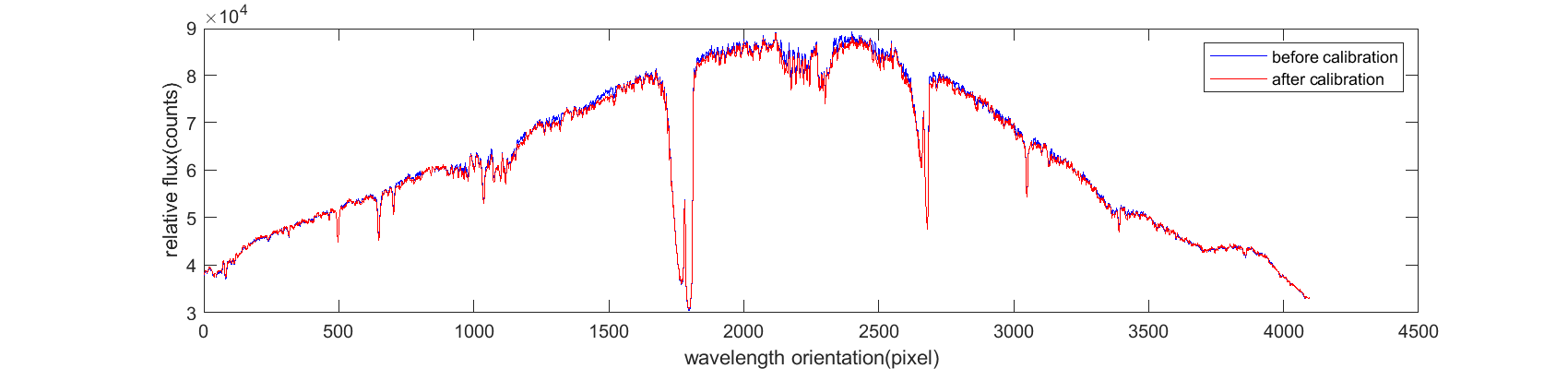}
    \caption{Flat-field spectrum extraction comparison before and after correction.}
    \label{fig:holeCompare}
\end{figure*}

\begin{figure*}
\label{fig:traceC}
\centering
\subfigure[Before aperture extraction]{
\begin{minipage}[t]{\linewidth}
\centering
\includegraphics[width=16cm]{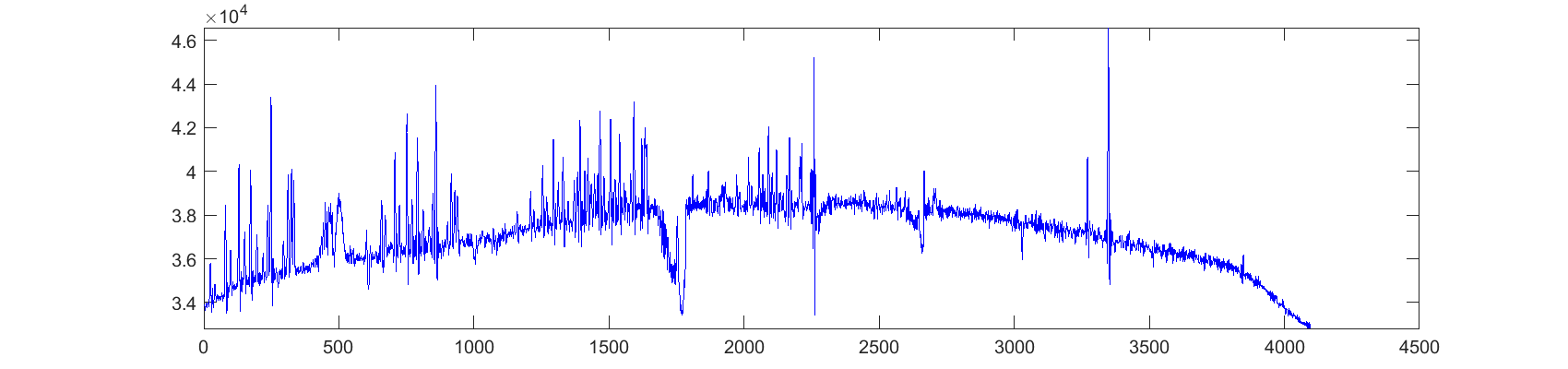}
\end{minipage}  %

}\\  %
\subfigure[After aperture extraction]{
\begin{minipage}[t]{\linewidth}
\centering
\includegraphics[width=16cm]{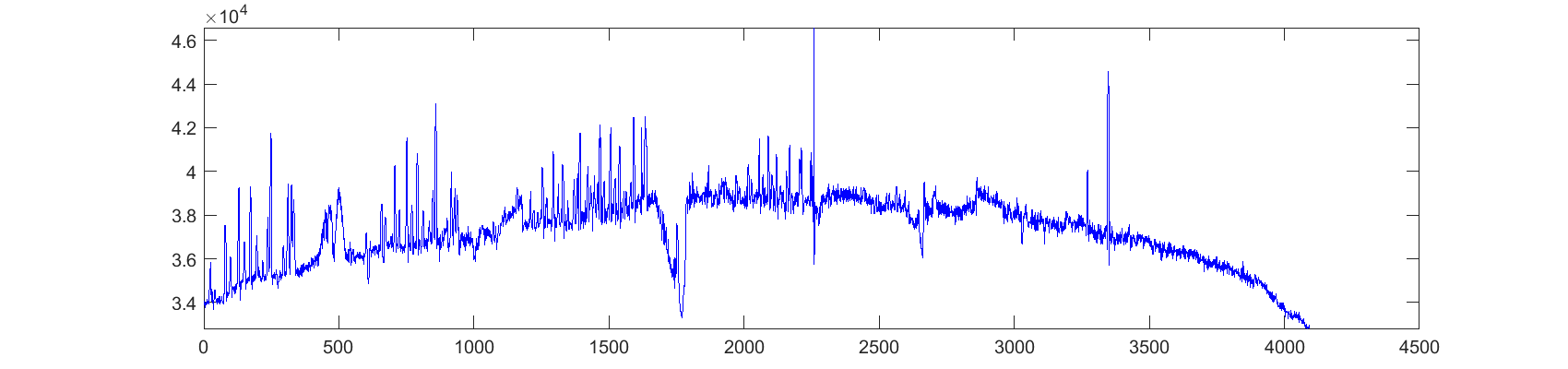}
\end{minipage}%
 
}%

\subfigure[Before blind deconvolution extraction]{
\begin{minipage}[t]{\linewidth}
\centering
\includegraphics[width=16cm]{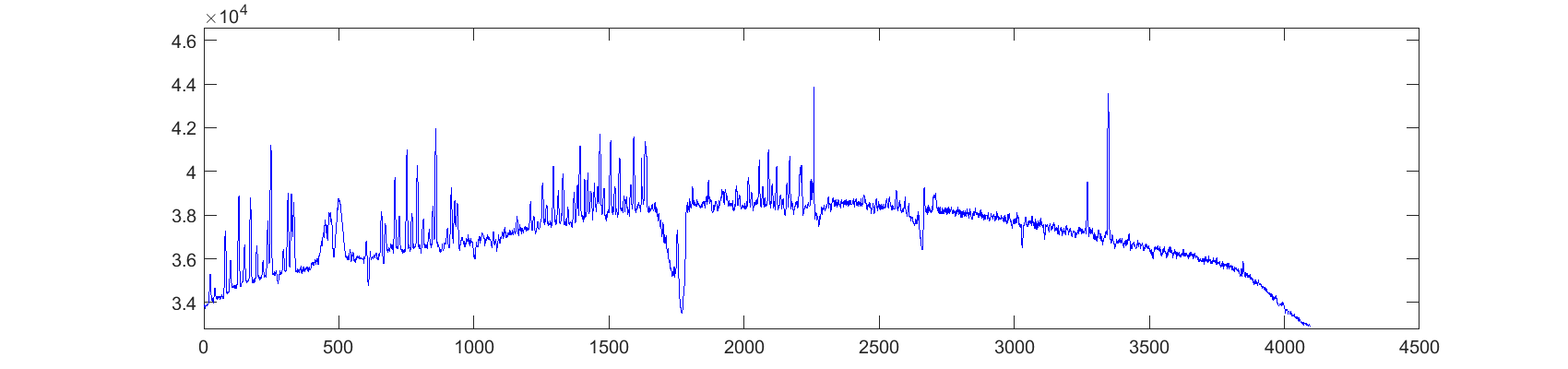}
\end{minipage}%

}%

\subfigure[ After blind deconvolution extraction]{
\begin{minipage}[t]{\linewidth}
\centering
\includegraphics[width=16cm]{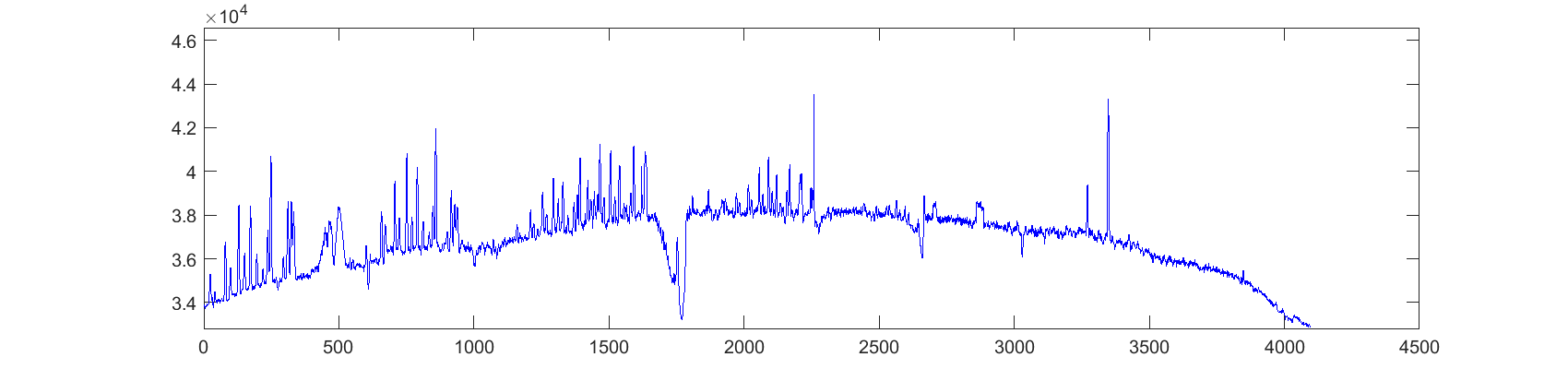}
\end{minipage}%

}%
\centering
\caption{Object spectrum extraction comparison before and after calibration.}
\end{figure*}

\begin{figure}
\centering
\subfigure[Before correction]{
\begin{minipage}[t]{0.3\linewidth}
\centering
\includegraphics[width=1in]{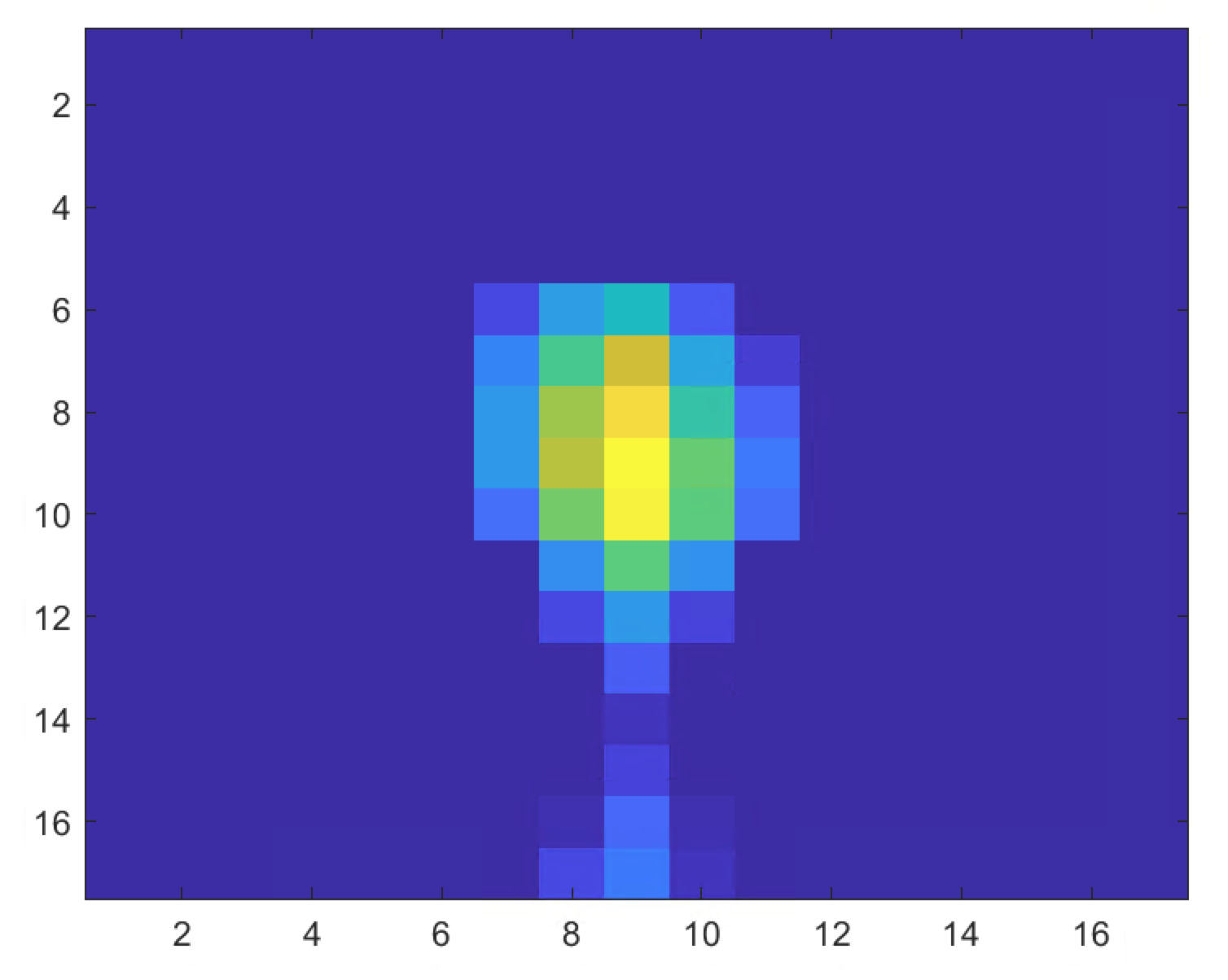}

\end{minipage}%
}%
\subfigure[After correction]{
\begin{minipage}[t]{0.3\linewidth}
\centering
\includegraphics[width=1in]{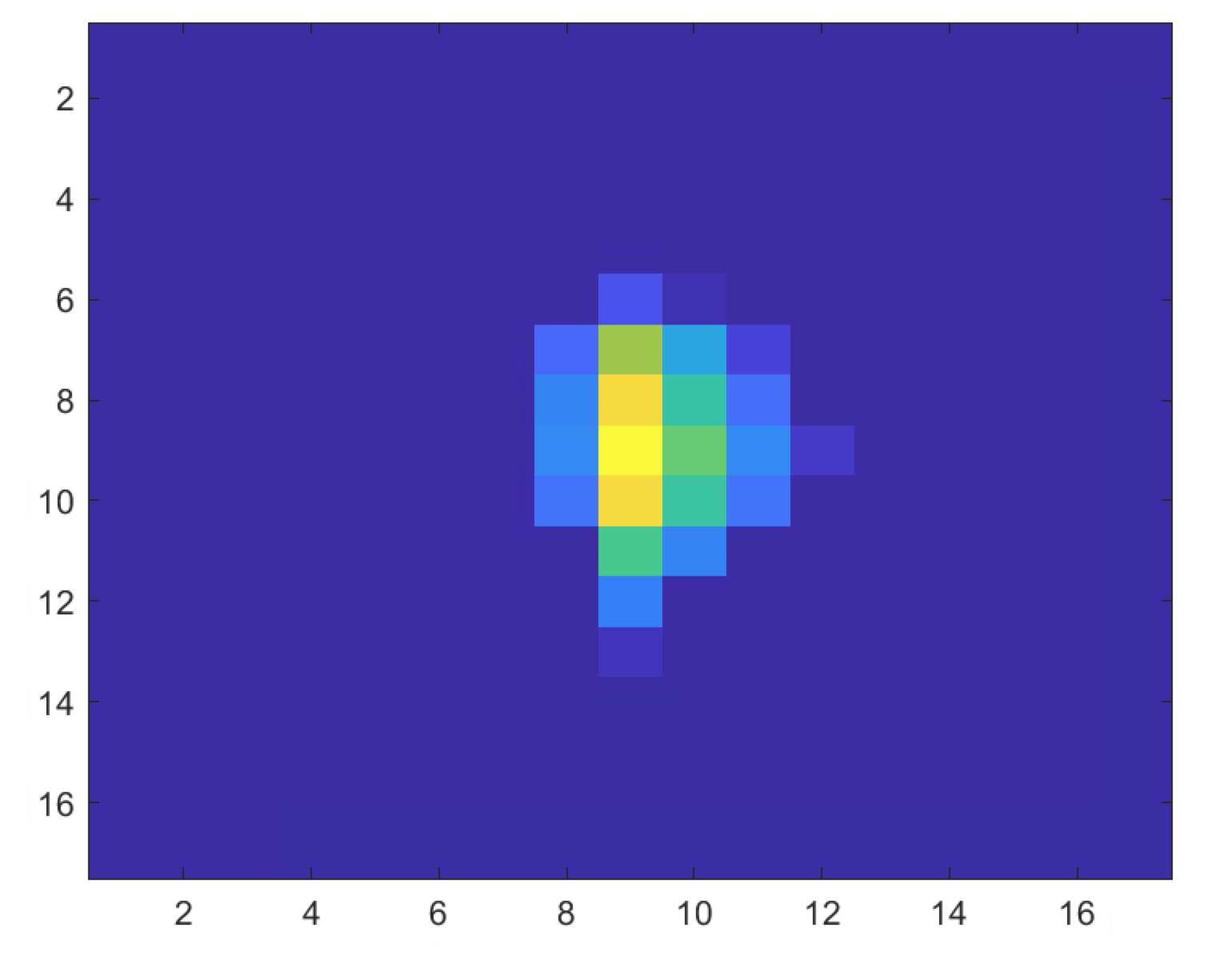}

\end{minipage}%
}%
\centering
\label{fig:psfCompare}
\caption{PSF comparison before and after correction.}
\label{fig:psfCom}
\end{figure}
 
 \begin{figure*}
    \centering
    \includegraphics[width = 0.6\linewidth]{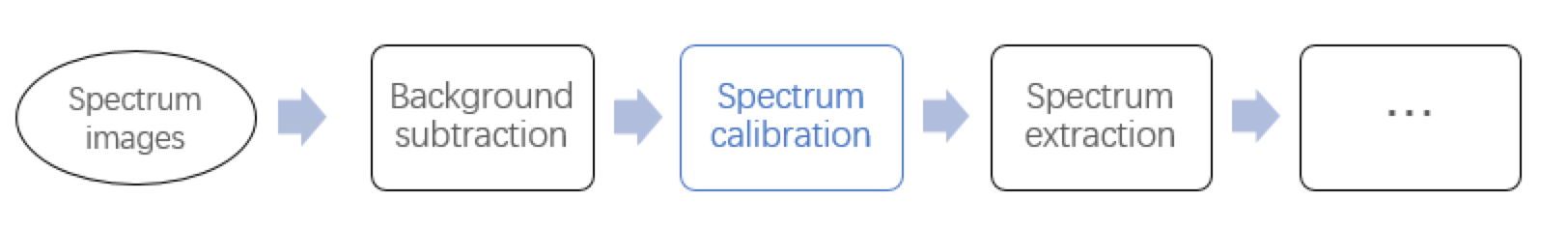}
    \caption{Data processing pipeline for  spectrum image correction.}
    \label{fig:newpipelines}
\end{figure*}

Due to the characteristics of the flat-field spectrum, we first correct the flat-field spectrum to show the effect of our method. Due to the large size of the flat-field spectral image and the dense arrangement between optical fibers, it is difficult to observe the full image with a resolution of $4096\times 4096$, so significantly distorted part with the resolution of $200 \times 200$ is enlarged to more clearly observe the effect of distortion calibration in Fig.\ref{fig:comp}, showing the comparison of the flat-field spectrum before and after the calibration, curved spectrum is obviously straightened.  Fig.\ref{fig:after_tracing} shows the fiber tracing result before and after the calibration, which further intuitively shows that the method proposed in this paper can effectively correct the distorted spectrum.

\par
Aperture extraction spectrum method is a particularly simple and practical method for extracting the 250 one-dimensional spectra from the two-dimensional spectral image, accumulating the flow value of all pixels along the spatial direction on both sides of the center of the spectrum with a certain radius. We performed spectrum extraction using this method to further analysis of the effects of spectrum calibration and  the spectrum extraction results before and after calibration are shown in Fig.\ref{fig:holeCompare}.
From the naked eye, the spectral characteristics before and after the calibration using aperture extraction remain basically the same, with a slight deviation in the wavelength direction and a significant increase in the flux at both ends, which meets our expectations. As we think, the curved spectrum is straightened, and the result of optical fiber tracing changes accordingly, so the result of the simple flux accumulation in spatial orientation will not very different, which at least shows that our proposed method does not cause a decline in the quality of the traditional spectrum extraction method. 
To demonstrate the effectiveness of the proposed method, we adopted a variance for evaluating the results of spectrum calibration. The variance \citep{wu2018var} of the extracted spectrum is calculated using equation~(\ref{eq:var}). 

\begin{equation}
    V=\frac{1}{N}\sum_{i=1}^{N}[spec(i)-\frac{1}{N} \sum_{N}^{j=1}spec(j)]^{2},
    \label{eq:var}
\end{equation}
where $N$ is the length of the spectrum and $spec(i)$ and $spec(j)$ denote the flux of the obtained spectrum. 

\par
The variance can reflect the smoothness of a spectrum, i.e., smaller variances signal smoother spectra, which indicate an effective spectrum calibration process. According to the calculation results, the result after correction is of less shake and smother. The focus of this paper is to utilize spectrum calibration to further improve the quality of blind deconvolution extraction, which is considered to be a more excellent method, the results of this part will be shown in Section~\ref{sec:object}.

\subsubsection{Object spectrum}

\begin{table*}
\caption{Variance of 20 flux lines of object spectra before and after calibration.}
\begin{tabular}{|l|l|l|l|l|l|l|l|l|l|l|}
\toprule  
                   & \multicolumn{10}{c|}{Variance (Fiber index)}                                             \\ \hline
Status             & 1      & 2      & 3      & 4      & 5      & 6      & 7      & 8      & 9      & 10     \\ \hline
Before & 0.0197 & 0.0175 & 0.0440 & 0.0596 & 0.0024 & 0.0184 & 0.0041 & 0.0084 & 0.0042 & 0.0072 \\ 
After   & 0.0136 & 0.0124 & 0.0391 & 0.0558 & 0.0014 & 0.0254 & 0.0039 & 0.0082 & 0.0041 & 0.0051 \\ 
\midrule  
                   & 11     & 12     & 13     & 14     & 15     & 16     & 17     & 18     & 19     & 20     \\ \hline
Before calibration &  0.0130 & 0.0079 & 0.0076 & 0.0272 & 0.0110 & 0.0058 &0.0189 &  0.0444& 0.0072 & 0.0135 \\ 
After calibration  &  0.0107 & 0.0054 & 0.0049 & 0.0235 & 0.0070 & 0.0056 & 0.0141 &  0.0375 & 0.0053 & 0.0102 \\ 
\bottomrule  
\end{tabular}
\label{tb:var}
\end{table*}

\begin{table*}
\caption{SNR of 20 flux lines of object spectra before and after calibration.}
\begin{tabular}{|l|l|l|l|l|l|l|l|l|l|l|}
\toprule  
                   & \multicolumn{10}{c|}{SNR(Fiber index)}                                             \\ \hline
Status             & 1      & 2      & 3      & 4      & 5      & 6      & 7      & 8      & 9      & 10     \\ \hline
Before & 91.73 & 38.20 & 31.28 & 59.34 & 58.05 & 35.46 & 78.79 & 47.50 & 61.75 & 117.05\\ 
After & 107.52 & 37.19 & 51.72 & 84.06 & 62.17 & 81.19 & 83.96 & 51.68 & 77.15 & 148.73 \\ 
\midrule  
                   & 11     & 12     & 13     & 14     & 15     & 16     & 17     & 18     & 19     & 20     \\ \hline
Before &  38.80 & 51.44 & 77.73 & 34.50 & 86.84 & 92.27 & 39.36 &  28.17& 66.87 & 46.86 \\ 
After  &  62.44 & 134.57 & 94.35 & 113.13 & 154.44 & 96.77 & 58.84 &  40.76 & 164.11 & 77.07 \\ 
\bottomrule  
\end{tabular}
\label{tb:snr}
\end{table*}

\label{sec:object}
We believe that the imaging distortion caused by the same spectrometer at a closer time is basically the same, so we also use the trained neural network model to correct the object spectrum captured by the same spectrometer on the same day. 
After calibration, two different methods were used to extract the same spectrum before and after calibration and show the effect, including aperture extraction and blind deconvolution extraction. The blind deconvolution extraction is our research work in 2017. Blind deconvolution is an image restoration technique in the case of unknown convolution kernel $f$. \cite{Bolton2010Spectro} proposed a simple cost formulation to be used for the blind deconvolution model and the algorithm is robust and high-performance. Through a large number of comparative experiments and analysis, \cite{Hang2017A} also verified that blind deconvolution could estimate stable PSF and obtain spectral results with robustness and high SNR. However, because the method does not take into account the spatially varying PSF, there are still some errors in final extracted result. In our proposed methodology, we mainly focus on improving the quality of spectrum extraction by blind deconvolution extraction through distortion calibration, which can make deconvolution method play a better role in practical applications.

In order to quantitatively evaluate the experimental results, we use the SNR and variance to measure the effect of spectrum extraction before and after calibration. The variances of the spectra extracted before and after calibration are both shown in Table~\ref{tb:var}, according to equation (\ref{eq:var}). The SNR \citep{QinFlux} is defined as,
\begin{equation}
   \text{SNR}=\frac{1}{N}\sum_{i=0}^{N-1}(F_{M}^{i}/\left | F_{i}-F_{M}^{i}\right |), 
\end{equation}
where $F_{i}$ is the flux of the extracted spectrum at the $i$th pixel, $F_{M}^{i}$ is the corresponding flux of the continuum, and $N$ is the number of selected pixels. The continuum is calculated by median filter with the width of 11 pixels. As there are massive sky lines in the entire spectra, the SNR is calculated by pixels in [0, 500] with a few sky lines.  The SNRs of the spectra extracted before and after calibration are both shown in Table~\ref{tb:snr} .

\par
Observing the results in the two tables, it can be clearly seen that the SNRs of the extracted spectra can be improved through spectrum calibration processing, and the one-dimensional spectral line variance obtained is smaller, indicating that the extracted spectra are more stable and smooth. Fig.\ref{fig:psfCom} shows the 2d figures of PSF estimated by blind deconvolution spectrum extraction before and after calibration. It is found that after image calibration, PSF changes significantly and presents a  centro- symmetric shape. This proves that the proposed two-dimensional optical fiber spectral image calibration method can eliminate the distortion in the imaging process. It is of great significance to extract one-dimensional spectra more accurately and less computationally intensively. Based on the above experimental results, we propose an improved two-dimensional spectrum processing flow shown in Fig.\ref{fig:newpipelines}. Spectrum calibration should be added to the current spectrum processing flow. Before  spectrum extraction, correcting the distorted spectrum can help improve the accuracy of the extraction, especially for the deconvolution method. On the one hand, the PSF of the points tends to be consistent, which is very helpful to reduce the calculation defects of deconvolution spectrum extraction. On the other hand, the spectrum calibration can help us to avoid the difficult problem of spatially changing PSFs modeling for large scale spectroscopy. Blind deconvolution method combined with spectrum calibration processing has better  extraction accuracy and higher application value in spectrum processing.



\section{Conclusions}
We propose a novel method  based on combined machine learning techniques  for two-dimensional optical fiber spectrum image correction. Firstly, utilizing  the prior knowledge of spectral imaging, we construct training samples from 250 tracing curves by fiber tracing for the need of neural network learning requirements. The input  of each training sample pair is the index value of the sampling pixel in both the wavelength and space orientations. Considering that neural network training is a kind of supervised learning, we adopt the  artificial transcendental knowledge and automatically label  unlabeled data according to the specific situation, constructing target output  of training sample pairs  by lines fitting. The corresponding output is the coordinate of the point after the TLS fitting. Compared with the Least Square method, TLS has considered the error of both independent and dependent variables, which improves the fitting accuracy of the tracing result. Secondly, combined with the fact that  less distortion in the centroid and noticeable distortion in periphery for  a spectrum image, we design a kind of non-uniform sampling technique  that projects the equidistant points on the arc, to select the samples efficiently and reduce computation complexity significantly of training neural network, at same time the feature extraction accuracy is not degraded.  The correction  experiments on  2D  flat-field spectrum and object spectrum data show that the correction  results of this method is very effective. At the same time, the spectrum extraction before and after correction was compared. The local PSF of the corrected object spectrum was estimated by blind deconvolution method, results illustrate that the PSF after correction is close to central symmetry, which indicates that this method has very significance in reducing the complexity  and improving accuracy of spectrum extraction.

\section*{Acknowledgements}
The research work described in this paper was fully supported by the Joint Research Fund in Astronomy (U1531242) under cooperative agreement between the NSFC and CAS, the grants from the National Natural Science Foundation of China(Project No.61472043). Prof. Qian Yin  and Prof. Ping Guo are the authors to whom all correspondence should be addressed.









\bsp	
\label{lastpage}
\end{document}